# Vernier Spectrum and Isospin State Control in Carbon Nanotube Quantum Dots

Jameson Berg[1*], Neda Lotfizadeh[1*], Dublin Nichols[2], Mitchell J. Senger[2], Wade De Gottardi[3†], Ethan D. Minot[2†] and Vikram V. Deshpande[1†]

[1]Department of Physics and Astronomy, University of Utah, Salt Lake City, Utah 84112, USA.

[2]Department of Physics, Oregon State University, Corvallis, OR 97331, USA.

[3]Department of Physics, Texas Tech University, Lubbock, TX 79409, USA.

[*]These authors contributed equally

[†]Correspondence to: wdegotta@ttu.edu, ethan.minot@oregonstate.edu, vdesh@physics.utah.edu

**Commensurability phenomena abound in nature and are typically associated with mismatched lengths, as can occur in quasiperiodic systems. However, not all commensuration effects are spatial in nature. In finite-sized Dirac systems, an intriguing example arises in tilted or warped Dirac cones wherein the degeneracy in the speed of right- and left-moving electrons within a given Dirac cone or valley is lifted. Bound states can be purely fast-moving or purely slow-moving, giving rise to incommensurate energy level spacings and a vernier spectrum. In this work, we present evidence for this vernier spectrum in Coulomb blockade measurements of ultraclean suspended carbon nanotube quantum dots. The addition-energy spectrum of the quantum dots reveals an energy-level structure that oscillates between aligned and misaligned energy levels. Our data suggest that the fast- and slow-moving bound states hybridize at certain gate voltages. Thus, gate-voltage tuning can select states with varying degrees of hybridization, suggesting numerous applications based on accessing this isospin-like degree of freedom.**

Commensuration [1] between scales is a ubiquitous phenomenon in physics and appears in areas as disparate as dislocations in epitaxial growth, moiré heterostructures, monolayer gas adsorbates, and superconducting vortex lattices. A daily-life example of commensuration is the vernier scale used in the laboratory to make accurate length measurements. Commensuration phenomena are not strictly spatial. For instance, vernier scales in frequency (so-called frequency combs) are used to perform vernier spectroscopy [2]. This work explores commensuration effects between energy levels in finite-sized quantum systems, i.e., quantum dots (QDs).

Carbon nanotubes (CNTs) are clean one-dimensional systems that exhibit quantum coherent transport and are actively investigated for the development of quantum technologies [3,4]. Potential applications are supported by progress in the production, separation, and alignment of CNTs [5]. The possibility of commensuration effects in the energy spectra of single-walled CNT QDs was first predicted by Izumida *et al.* [6,7]. Such intrinsic electrical commensuration is connected to a host of other topics in contemporary CNT research [7–12].

Despite a significant body of experimental work on quantum bound states in CNTs, previous experiments have not observed a vernier spectrum of energy levels. Single-electron energy levels in CNT QDs are typically measured by electron transport experiments. The physics that is accessible in these experiments depends on (i) the length of the CNT, (ii) the environment surrounding the CNT, and (iii) the nature of the electrical contacts. Long CNTs can hold more electrons, which is needed to measure the energy-level spacing for hundreds of electrons. Ultra-clean CNTs are required so electronic disorder does not disrupt patterns in the energy-level spacing. The electrical contacts must be either opaque or semi-transparent so that single-electron charging effects (Coulomb blockade) can facilitate the measurement of shell filling. Previous measurements of shell filling used short CNTs (a few hundred nanometers) [13,14]. Previous reports of long CNT (> 1 μm) in ultra-clean environments (fully suspended) have used transparent contacts [15–17]. Our current work is unique because we study ultra-clean suspended CNT with semi-transparent contacts and length > 2 μm (see Fig 1a for a device schematic).

In electron transport measurements of quantum dots, a gate voltage, $V_g$, is used to tune the number of electrons, $N$, that occupy the quantum dot. Transitions in occupation number (from $N$ to $N + 1$) are detected as a peak in zero-bias conductivity, as shown in Fig 1c and 1d (Coulomb blockade peaks). If the electron energy levels of the QD have 4-fold degeneracy, the Coulomb blockade peaks form groups of four (Fig. 1c). If energy levels have 2-fold degeneracy, the Coulomb peaks form groups of two (Fig. 1d). Further details about the energy-level structure can be extracted by measuring $\Delta V_g$, the potential between consecutive Coulomb peaks. The potential $\Delta V_g$ is proportional to the energy required to add the $N^{th}$ electron to the quantum dot (see the addition energy plots in Fig 1c and 1d). Previous literature on CNT QDs has sometimes reported 4-fold shell filling [14,18] and sometimes reported 2-fold shell filling [13,19]. There have been no previous reports of a systematic oscillation between 4-fold and 2-fold shell filling.

In our current work, we measure the addition energies of hundreds of electrons on long, ultra-clean CNT QDs. We find oscillating patterns in the addition energies, matching the vernier predictions of Izumida et al. We also find new features that were not previously predicted by theory. For example, some vernier lobes are rounded, in contrast to the sharply peaked structures predicted by standard theory; moreover, adjacent lobes alternate in height. We extend existing theory to show that our experiment probes the degree of polarization/hybridization of the various quantum states probed in the system.

Figure 1b illustrates key features in the band structure of metallic CNTs, which are expected to produce a vernier spectrum. There are two valleys centered at the K and K' points. Each valley hosts left-moving and right-moving electrons. The simplest model of CNT band structure predicts linear bands with symmetry between left-/right- moving states [19]. However, higher-order corrections such as trigonal warping [15–17] give rise to non-linear bands and asymmetric velocities for left-/right- moving states. The velocity asymmetry was recently demonstrated in electron transport experiments on ultra-long CNTs with transparent electrical contacts [15,16]. In the case of opaque or semi-transparent contacts (as studied here), the CNT hosts electronic bound states with well-defined energy levels. Intervalley scattering at either end of the CNT couples a fast left-moving state in the K valley to a fast right-moving state of the K' valley (following the notation introduced by Izumida et al., we use $\tau_z$ to denote this band degree of freedom; see the arrow labeled as $\tau_z = +1/2$ in Fig. 1b). Similarly, a slow right-moving state of the K valley is coupled to a left-moving state of the K' valley (see the arrow labeled as $\tau_z = -1/2$ in Fig. 1b). For energies below the charge neutrality point (for holes), the level spacing between adjacent $\tau_z = +1/2$ states are slightly larger than for $\tau_z = -1/2$ states.

Figure 1e illustrates the energy level diagram resulting from a 14% mismatch in the level spacing of fast-moving versus slow-moving states. When the fast- and slow-moving states align in energy (highlighted in green), transport measurements would detect 4-fold shell filling. When the fast- and slow-moving states are maximally misaligned (highlighted in pink), transport measurements would detect 2-fold shell filling. Figure 1f illustrates the addition energy spectra associated with Fig 1e. There is an oscillation between regions of 4-fold shell (highlighted in green) and regions of the 2-fold shell filling (highlighted in pink). To visualize the structure of the oscillations, we color the addition of the $N^{th}$ electron according to the value of $N$ mod 4. The case of *N* odd is indicated by white dots, which form a flat baseline corresponding to the charging energy of the quantum dot. The case of *N* even is indicated by black (*N* mod 4 = 0) and grey (*N* mod 4 = 2) dots, which form interlacing tent-shaped structures, referred to here as vernier lobes.

Our measurements of suspended CNT quantum dots with length > 2 µm were acquired at 1.5 K using a variable-temperature pumped helium cryostat. The CNTs were grown directly over the trenches via chemical vapor deposition using techniques detailed in previous works [20,21]. Figure 2 shows three different tubes that all exhibit remarkably clean shell-filling Coulomb oscillations (see insets of each tube) over a very large gate voltage range. The shell-filling oscillations are on top of long-period Sagnac oscillation (quantum interference effect) that results from the asymmetric velocities of charge carriers in the different bands [15,16]. Electron addition energies were extracted by measuring the gate voltage spacing between the maximum of each shell-filling peak. To ensure the true peak was extracted, a cubic spline interpolation was performed on the conductance data sets, the details of which are explained in S.2. Further we used the coloring scheme for electron number modulo four as explained in Fig 1f to reveal any sub-structure. The resulting color-coded addition energies and corresponding conductance plots are shown in Fig. 2. Clear vernier spectra are seen in all three tubes over a large range of gate voltage with interlacing black and grey dots. This pattern is present in every tube measured in the shell-filling regime (over 20 tubes), with the only exception being tubes where the level spacing is comparable to the thermal energy.

The vernier shell-filling is characterized by various lobes. By comparing the theoretical shell-filling patterns in Figure 1f with the data in Fig. 2, we see that the nodes of the vernier lobes (crossing of black and grey dots) occur when the two bands are maximally separated in energy, which is associated with period 2 patterns of the addition energies (Fig 1d). The antinodes (maximum separation between black and grey) occur when the two bands have nearly degenerate energies, corresponding to a period 4 pattern of the addition energies (Fig 1c). Another notable feature in the signature is the lack of a charging energy floor. In each tube, the charging energy floor, denoted by white circles, varies with the chemical potential. The vernier lobes in the addition energy spectrum reveal another pattern: the vernier lobes alternate in size, i.e., the peaks consisting of grey dots are a few meV higher than the peaks consisting of black dots. Notably, in Device 3 (Fig. 2c), the alternate lobe peaks are suppressed to such a degree that they are no longer visible at $V_g$ = -3.7 V. Such a pattern in lobe height is not predicted in theories of the vernier spectrum in CNTs [6,7].

To address the origin of the alternating vernier lobe patterns and the implications of the vernier energy spectrum in general, we consider a phenomenological model that takes single-electron states that are pure $\tau_z$ states as its starting point. The rounding of the vernier lobes seen in the data (Fig. 2) is suggestive of level repulsion, indicating that a perturbation is mixing the $\tau_z$ bands. Numerous effects can and likely do contribute to this rounding, including spin-orbit coupling [22] and Coulomb interactions [23]. In supplementary Figs. 3d-e, we discuss the effect of these factors on the charging energy floor. However, neither of these effects accounts for the conspicuous pattern of alternating heights of the vernier lobes seen in Fig. 2.

The even/odd pattern in Fig. 2 can be accounted for by considering the effect of a non-uniform potential along the tube. As justified below, we take a potential that is sharply peaked near the ends of the tube

$$V(x) = V_0 a[\delta(x) + \delta(x - L)], \tag{1}$$

where $V_0$ and $a$ are phenomenological parameters with dimensions of energy and length, respectively. A theoretical shell-filling spectrum is calculated by diagonalizing Eq. (1) for the single electron states whose wave functions are $\psi_\pm(x) = \cos(\pi n_\pm x/L)/\sqrt{2L}$, where $n_\pm$ is the number of electrons in the $\tau_z = \pm 1/2$ bands, respectively (see Supplementary Information). The shell-filling exhibits sharply-peaked vernier lobes for the case that $V_0 = 0$, shown in Fig. 3a. For finite $V_0$, level-repulsion gives rise to rounding of the vernier lobes and the alternating pattern associated with adjacent vernier lobes. The latter arises due to the form of the potential (1) and the relative parities of the nearly degenerate states in different lobes. The parity of the state $\psi_\pm(x)$ is $P = (-1)^{n_\pm}$ so that within a given band, the parities of the states alternate with energy. Given that the potential (1) is even about $x = L/2$, the matrix element between states of opposite parity vanishes, while it is non-zero between states of the same parity. In going from one vernier lobe to the next, the number of states added with $\tau_z = +1/2$ and $\tau_z = -1/2$ differs by one and so the relative parity of the nearly degenerate states switches from one lobe to the next. As a result, if the matrix elements of $V(x)$ are zero between the nearly degenerate states in one lobe, they are non-zero in the next. The effect of level repulsion alternates in strength, which is broadly consistent with the data. We note that this alternating pattern is inconsistent with an impurity potential somewhere in the middle of the tube, which would generally lead to an aperiodic pattern of envelope heights. Theoretical addition energies for a tube with parameters $V_0 = 0.3$ meV and $a = 0.1\,L$ are plotted in Fig 3b, where the extra even/odd vernier lobe pattern is now present and resembles that of Fig 2. The form of potential (1) suggests that the leads at the tube's ends are responsible for this effect, at least in part.

As a result of the coupling (1), the single-electron states are no longer pure eigenstates of $\tau_z$ but instead are hybridized. In order to quantify this, we have evaluated the expectation values $\langle \tau_z \rangle$ for the states in the simulated shell-filling spectra shown in Figs 3a and b. The hybridization pattern closely tracks the vernier oscillations, with the nearly degenerate states (Fig. 3b) in every other vernier lobe being maximally hybridized marked by arrows colored with the $\tau_z$ hybridization. Most other states are nearly pure band eigenstates with $\langle\tau_z\rangle \approx \pm 1/2$ (Fig 3a). Given the regularity of this pattern, we may estimate the degree of hybridization for actual data (Fig. 2) by comparing it with these theoretical predictions. The predicted degrees of hybridization are overlaid on the experimentally measured addition energies of Device 1 in Fig. 3c. It is clear that the gate is able to tune the band quantum states from maximally polarized to hybridized. Such gate control over an isospin degree of freedom may be particularly useful in the study of valleytronics in the increasingly prevalent mixed-dimensional van der Waals heterostructures [24–26]. For example, in 1D-2D hybrid systems made from CNTs and graphene or transition metal dichalcogenides (TMDs), it is predicted that valley coupling between the 1D and 2D systems with idealized band structures (i.e., without left/right asymmetry) leads to strong correlations and similar asymmetries and valley polarization/hybridization [27] as observed in our work.

In conclusion, the vernier energy scale arises in finite-sized CNTs when the two bands have slightly detuned energy level spacings. The addition energies of CNTs in the shell filling regime produce characteristic gate dependent vernier lobes over a large $V_g$ range, from the presence of a vernier scale in the isospin energy levels. As a result of the quality of the transport data of our ultra-clean suspended CNTs, new features of the vernier spectrum have been revealed. These vernier lobes have an alternating envelope height, which points to the coupling of the band isospin degree of freedom by an external potential associated with the ends of the nanotubes. We have used a phenomenological model to estimate the degree of isospin polarization for each of the energy states. This suggests that CNTs can serve as a rich platform where states of any desired polarization can be selected using the gate voltage. Taken together with the recent progress in the fabrication of hybrid mixed-dimensional systems [24–26] using CNTs and graphene or TMDs, our work suggests the possibility of using CNTs as a source for quantum states whose isospin degree of freedom can be electrically tuned. This work has clear implications for valleytronic circuits.

**Acknowledgments**

Work performed in Utah was supported in part by the National Science Foundation under Grant No. 2005182. Work performed in Oregon was supported by the National Science Foundation under Grant No. 1709800. A portion of device fabrication was carried out in the University of California Santa Barbara (UCSB) nanofabrication facility. Part of this research was conducted at the Northwest Nanotechnology Infrastructure, a National Nanotechnology Coordinated Infrastructure site at Oregon State University which is supported in part by the National Science Foundation (grant ECCS-1542101) and Oregon State University.

**Competing interests**

The authors declare no competing interests.

**Figures**

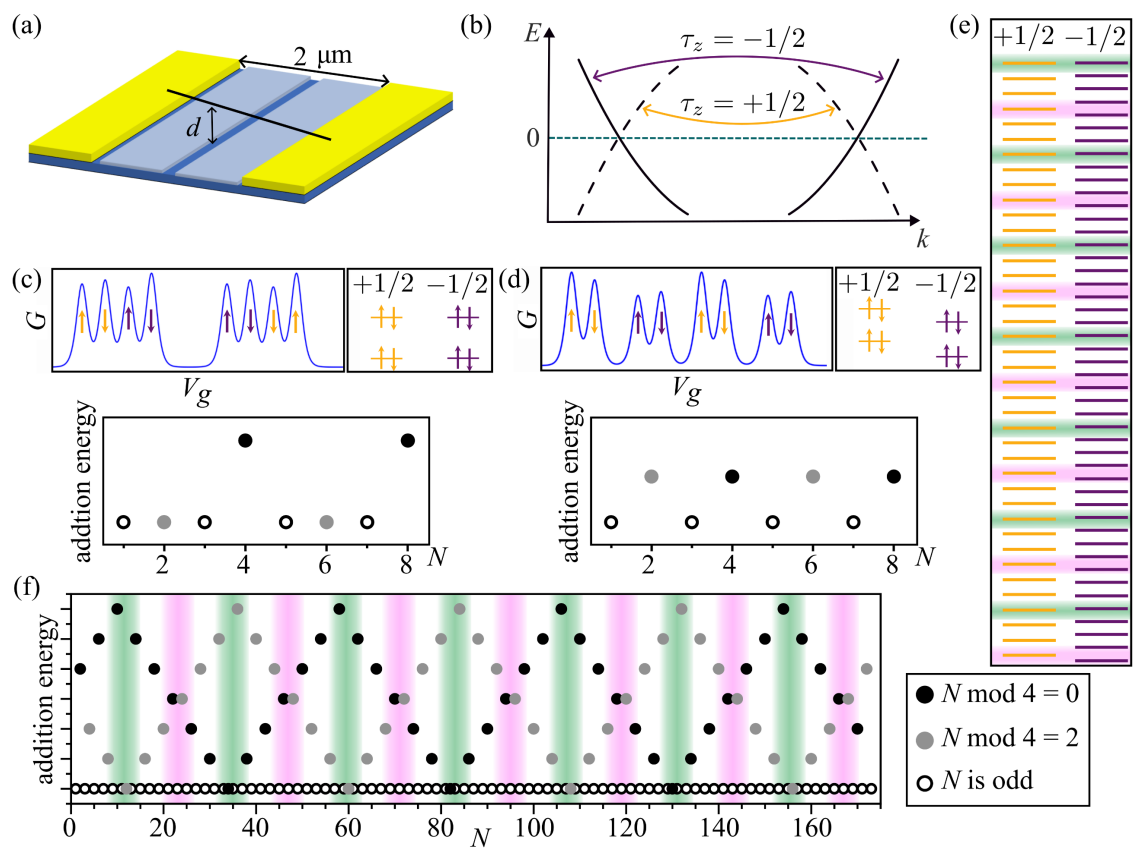

**Figure 1. (a)** Schematic diagram of the suspended CNT quantum dot, with two gate electrodes under the CNT. Both gates are set to the same voltage, $V_g$. **(b)** Sketch showing the composition of band-polarized state, which are formed from a superposition of a right- (left-) mover from one valley and a left- (right-) mover from the other valley. The band isospin is indicated by the quantum number $\tau_z$ = +1/2 and –1/2. **(c-d)** Sketch of conductance vs $V_g$ (top panel) and addition energy vs electron number, $N$, (bottom panel) for tubes in the quantum dot regime showing **(c)** four- and **(d)** two-electron periods, respectively. **(e)** For energies less than the charge neutrality, the energy level spacing associated with the +1/2 bands are greater than those in the –1/2 band. The mismatch in energy level spacing gives rise to the vernier shell-filling spectrum. **(f)** The addition energy vs. electron number, $N$, for a CNT with a vernier energy spectrum exhibits characteristic peaks created by the transition between four- and two-period states. (Energies with 4-electron periods noted in green and points of two-electron periods noted in pink.)

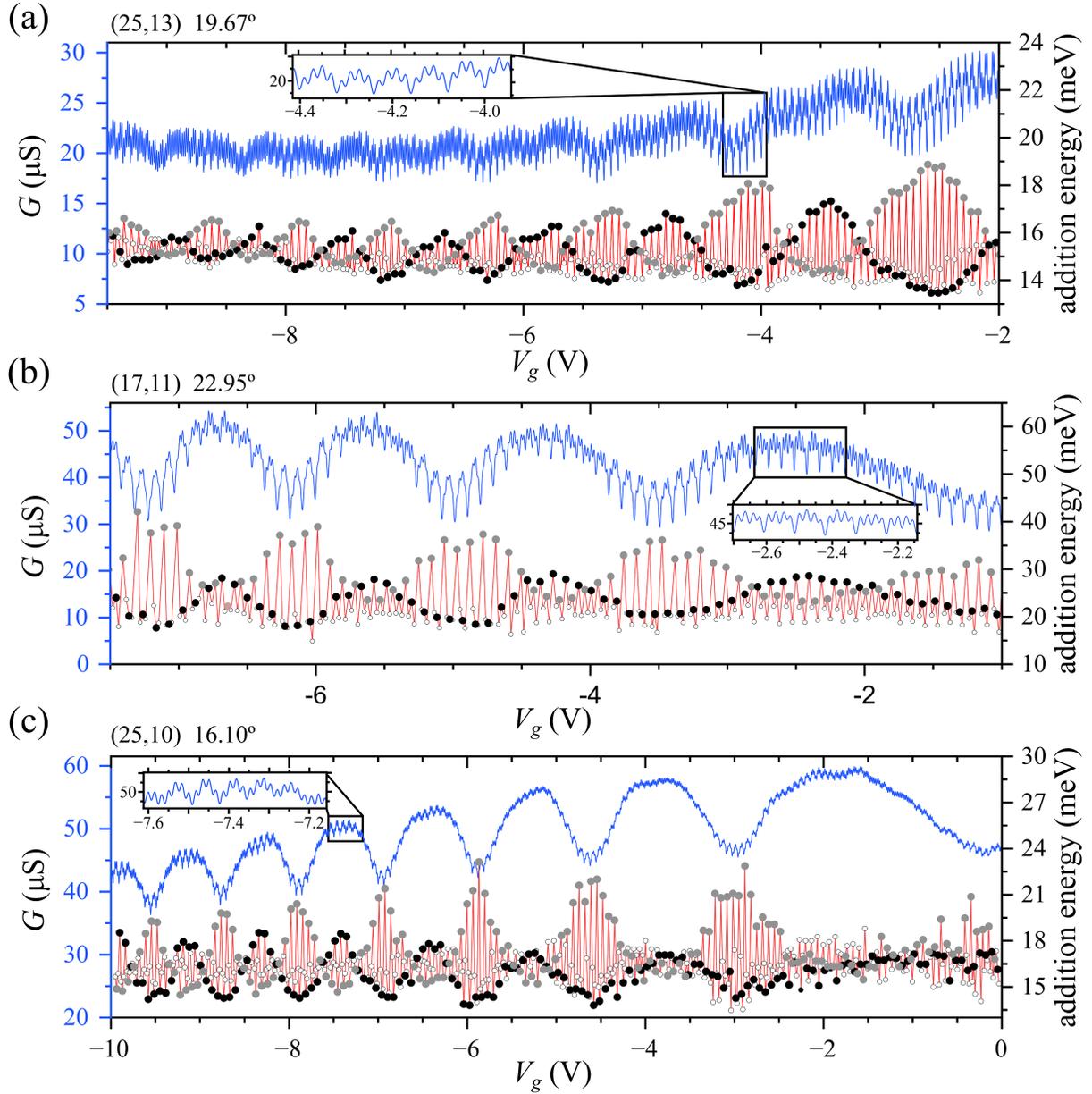

**Fig. 2. (a-c)** Measured conductance, $G$, versus $V_g$ plotted in blue for Devices 1, 2 and 3 in the hole doped regime (negative $V_g$) at $T$=1.5 K. Each device has clean shell filling peaks for over more than 400 electrons. The insets highlight representative clean shell filling peaks. Plotted in red is the corresponding addition energy for each electron added to the CNT, with $N$ odd electrons marked with a white circle and $N$ even electrons indicated by black ($N$ mod 4 = 0) and grey ($N$ mod 4 = 2) dots. All three devices show characteristic vernier lobes in the addition energy graph. The vernier lobes alternate in size in an even/odd pattern. Measured chiral indices and angles of each tube are noted in the top left corner of each plot.

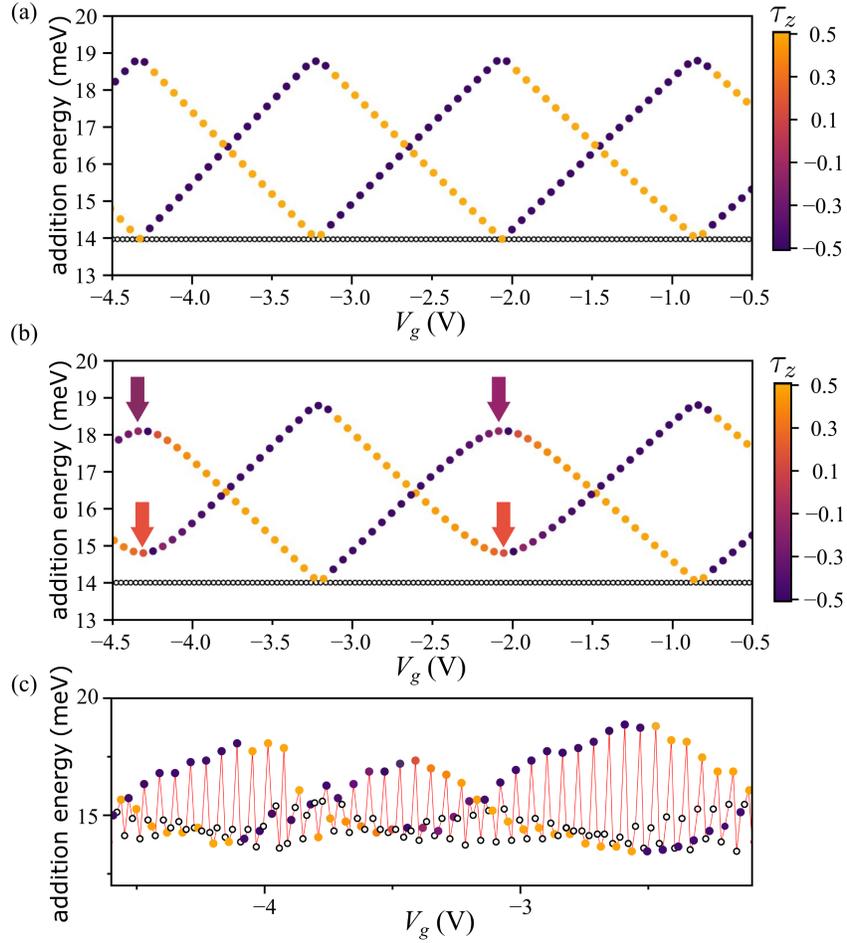

**Fig. 3. (a)** Calculated shell filling structure with no lead coupling ($V_0 = 0$) for a 2.5 μm tube with chiral indices (25,10). The electron number is with respect to half-filling. **(b)** Same as (a) with coupling turned on ($V_0 = 0.3$ meV and $a = 0.1\,L$). Color scheme shows degree of hybridization of states between $\tau_z = +1/2$ and $\tau_z = -1/2$ and odd electrons denoted by white circles. In vernier lobes with reduced peak heights, $\tau_z$ states become hybridized with arrows showing the points of maximal hybridization. **(c)** Section of addition energy from Fig 2a with hybridization marked following the patterns present in the model.